\documentclass[aps,prb,twocolumn,groupedaddress,showpacs]{revtex4}

\RequirePackage{graphicx}

\begin{document}


\title{Quantum magnetism of perfect spin tetrahedra in Co$_{4}$B$_{6}$O$_{13}$}


\author{Hiroyuki Hagiwara}
\author {Hirohiko Sato}
\email[]{hirohiko@phys.chuo-u.ac.jp}
\affiliation{Department of Physics, Chuo University, 1-13-27 Kasuga, Bunkyo-ku,
Tokyo 112-8551, Japan.}
\author{Masahiro Iwaki}
\author{Yasuo Narumi}
\altaffiliation[Present address: ]{Institute for Materials Research, Tohoku University, 2-1-1 Katahira, Aoba-ku, Sendai 980-8577, Japan}
\author{Koichi Kindo}
\affiliation{The Institute for Solid State Physics, The University of Tokyo, 
5-1-5 Kashiwanoha, Kashiwa 277-8581, Japan}

\date{\today}

\begin{abstract}
Co$_{4}$B$_{6}$O$_{13}$ contains undistorted tetrahedral clusters of magnetic Co$^{2+}$ ions. The high-field magnetization of this magnet exhibits a periodic undulation indicating quantization of the total spin number per cluster. Measurements of magnetic susceptibility and specific heat reveal that the ground state is composed of several different singlet states, reflecting the high symmetry of the spin tetrahedron. An exact diagonalization calculation taking account of single-ion type anisotropies and Dzyaloshinsky-Moriya interactions reproduces the expretimental results very well.
\end{abstract}


\maketitle


\section{Introduction}

Spin, the elementary unit responsible for magnetism, is essentially a quantum mechanical property.\cite{messiah99} The most stable state for an antiferromagnetic spin pair (dimer) is the spinless singlet state. As the wave function of the singlet state is an entanglement of up-down and down-up spin states, there is no classical analogue. Nevertheless, most antiferromagnets, in which the magnetic ions form a uniform three-dimensional lattice, can be well described by a classical (trivial) model of an alternating arrangement of antiparallel spins.\cite{blundell01} Systems in which geometrical constraints prevent the formation of such a trivial structure have received considerable attention in recent years. In such systems, known as frustrated magnets, quantum behaviors are particularly enhanced.\cite{ramirez94} The simplest example of a frustrated system is a triangle of three spins that interact antiferromagnetically. This configuration clearly precludes stable resolution of all three antiferromagnetic interactions. This situation is exacerbated in a spin tetrahedron. 

A three-dimensional network of vertex-sharing tetrahedra constitutes an extremely frustrated system when the nearest interactions are antiferromagnetic. Such a lattice is known as a pyrochlore lattice. As pylochlore-lattice magnets have a vast number of microscopic states of very similar energies, the magnetic phase is very sensitive to subtle perturbations that can be otherwise ignored in non-frustrated systems. Antiferromagnetic pyrochlore-lattice magnets, such as ZnCr$_{2}$O$_{4}$, CdCr$_{2}$O$_{4}$ and HgCr$_{2}$O$_{4}$, display both lattice distortion and long-range magnetic ordering.\cite{lee00,matsuda07,ueda05} In this kind of materials, the true ground state of the undistorted pyrochlore lattice is usually hidden by the effect of lattice distortion, which relieves the high frustration and imparts a long-range ordered state. On the other hand, the pyrochlore antiferromagnet Y$_{2}$Mo$_{2}$O$_{7}$ has been shown to undergo a transition to a spin-glass state\cite{gingras97,gardner99} in which spins are frozen without the formation of long-range order. The spin glass can be well described within a classical mechanical framework and the true ground state is still hidden.

Theoretically, the precise ground state of pyrochlore lattices with quantum ($s = 1/2$) spin, however, remains controversial, with some theories predicting a spin-liquid state and others predicting spin-singlet ordering.\cite{canals98,tsunetsugu01,koga01} The concept of a spin liquid has been proposed for a triangular lattice of $s = 1/2$ spins, in which all spins form singlet pairs but the combination of pairs fluctuates.\cite{anderson73,shimizu03} However, it has been pointed out that the pyrochlore lattice contains not only dimer singlets, but also other types of singlet states, which have been referred to as plaquette singlets.\cite{koga01} The coexistence of different types of singlet states is expected to have an important effect on the ground state of pyrochlore magnets.

Apart from such infinite systems, direct experimental evidence for a simpler system composed of almost isolated tetrahedra is expected to provide valuable information clarifying the relationship between highly symmetric structures and the quantum aspects of spins. The aim of the present study was to characterize the behavior of a perfect spin tetrahedron formed by four magnetic ions without lattice distortion. However, lattices comprised of perfect spin tetrahedra are exceedingly rare, as the spin tetrahedra in most model materials are distorted.\cite{schnack06} Recently, the present authors succeeded in synthesizing good single crystals of Co$_{4}$B$_{6}$O$_{13}$\cite{rowsell03} which is one of the ideal materials for studying a spin tetrahedron. We report here pronounced quantum behaviors of its magnetism. 

\section{Experimental}
Single crystals of Co$_{4}$B$_{6}$O$_{13}$ are synthesized using a hydrothermal method. A mixture of Co(OH)$_{2}$ and B$_{2}$O$_{3}$ are sealed in a silver capsule with NaOH solution and then it was heated up to 630$^{\circ}$C under a 150 MPa of hydrostatic pressure in a test-tube type autoclave. After the reaction for a few days, dark purple crystals with truncated-cubic shape are obtained. The volume of a typical crystal is 1-10 mm$^{3}$, large enough for the measurements of magnetic and thermal measurement. We confirmed that each of the obtained crystals has the lattice constant of Co$_{4}$B$_{6}$O$_{13}$, $a=7.4825$~\AA,  reported by Rowsell, \emph{et al.}\cite{rowsell03} using an imaging-plate type x-ray diffractmeter (Rigaku, R-Axis Rapid). The magnetic susceptibility between 2 K and 300 K was measured using a superconducting-quantum-interference-device (SQUID) magnetometer (Quantum Design MPMS-XL) at a 1 T of matnetic field. The specific heat was measured down to 0.4 K on a single crystal using a commercial instrument (Quantum Design, PPMS). The magnetization curves up to 70 T at 4.2 K and 1.3 K were measured using a hand-maded pulse magnet installed at the Institute for Solid State Physics.

\section{Results and Discussion}
\begin{figure}[tb]
\begin{center}
\includegraphics[width=1.0\linewidth]{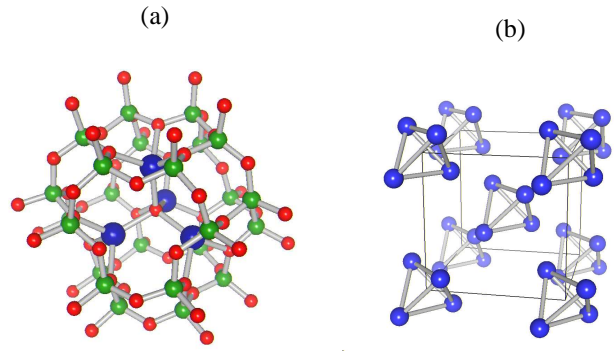}
\end{center}
\caption{Crystal structures of Co$_{4}$B$_{6}$O$_{13}$. (a) Coordination environment of Co atoms within a Co$_{4}$ cluster (large, Co; medium, B; small, O). Adjacent cobalt ions are bridged by an oxygen atom located at the center of the tetrahedron. (b) Crystal structure showing only Co atoms. The Co$_{4}$ clusters form a body-centerd-cubic lattice by connection of uniformly oriented clusters via a bulky boroxide network.}
\label{fig1}
\end{figure}

The crystal lattice of Co$_{4}$B$_{6}$O$_{13}$ contains regular tetrahedra formed by four Co$^{2+}$ ions (Fig.~\ref{fig1}(a)). Within these Co$_{4}$ clusters, adjacent cobalt ions are bridged by an oxygen atom located at the center of the tetrahedron with an intracluster Co-Co distance of 3.24 \AA. The Co$_{4}$ clusters further form a body-centered-cubic lattice (Fig.~\ref{fig1}(b)) by connection of uniformly oriented clusters via a bulky boroxide network with an intercluster Co-Co distance of up to 4.26 \AA. The space group of this crystal lattice is $I\bar{4}3m$ and the crystallographic sites of Co$^{2+}$ ions are unique. The uniform orientation of tetrahedra makes it possible to measure the macroscopic anisotropic properties of the Co$_{4}$ tetrahedra. The tetrahedra in Co$_{4}$B$_{6}$O$_{13}$ were confirmed to be undistorted down to a temperature of 0.4 K through a combination of powder X-ray diffraction measurements down to 8 K and specific heat measurements down to 0.4 K. No structural phase transitions were apparent in any of these measurements.

\begin{figure}[tb]
\begin{center}
\includegraphics[width=1.0\linewidth]{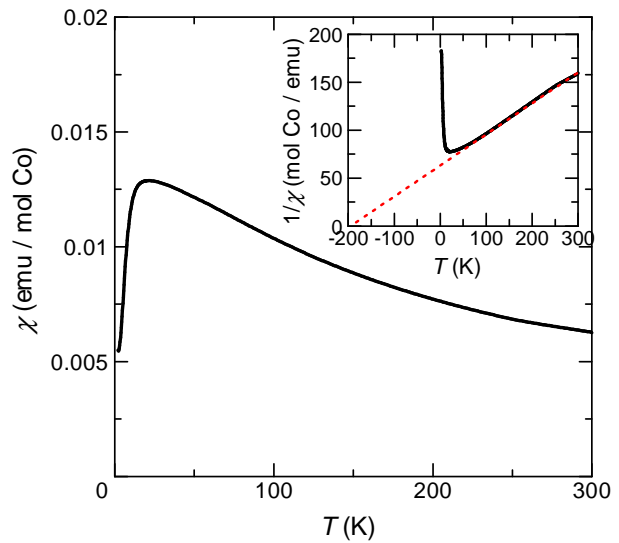}
\end{center}
\caption{Temperature dependence of magnetic susceptibility of Co$_{4}$B$_{6}$O$_{13}$ measured down to 2 K under a 1 T magnetic field applied parallel to the [111] direction. The inset shows the reciprocal susceptibility demonstrating Curie-Weiss behavior at high temperature above 100 K.}
\label{fig2}
\end{figure}

The magnetic susceptibility of Co$_{4}$B$_{6}$O$_{13}$ measured under a 1 T magnetic field at temperatures down to 2 K is shown in Fig.~\ref{fig2}. At high temperature above 100 K, the susceptibility is well reproduced by the Curie-Weiss function $\chi = C/(T - \Theta)$, where $C$ is the Curie constant and $\Theta$ is the Weiss temperature, respectively. For Co$_{4}$B$_{6}$O$_{13}$, $C = 2.98$ emu$\cdot$K/mol~Co, and the effective magnetic moment ($\mu_{\mathrm{eff}}$) is estimated to be 4.90 $\mu_{\mathrm{B}}$. Taking account of the high-spin ($S = 3/2$) state of Co$^{2+}$, the $g$-factor for Co$_{4}$B$_{6}$O$_{13}$ is calculated to be 2.53, which is reasonable for tetrahedrally coordinated Co$^{2+}$ ions.\cite{sturge69} The large absolute value of Weiss temperature ($|\Theta|$ = 190 K) is indicative of very strong antiferromagnetic interactions. In the low-temperature range, the susceptibility exhibits a broad maximum at 14 K, below which the susceptibility drops to a small value. This behavior differs from that typical of an antiferromagnetic transition in that there is a characteristic sharp bend in the susceptibility/temperature curve, instead undergoing a crossover to almost non-magnetic ground state at low temperature.\cite{kageyama99}

\begin{figure}[tb]
\begin{center}
\includegraphics[width=1.0\linewidth]{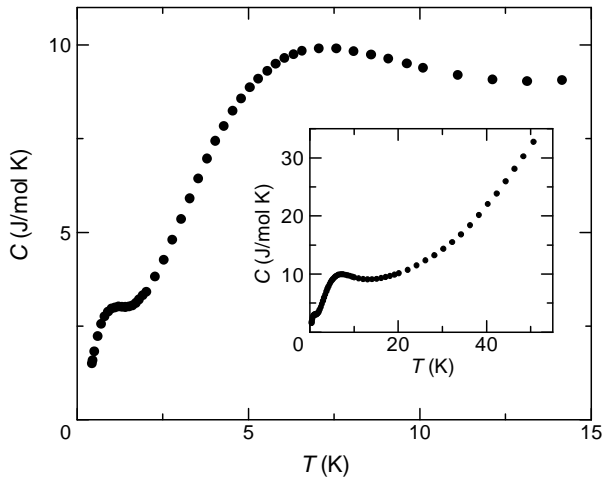}
\end{center}
\caption{Temperature dependence of specific heat of Co$_{4}$B$_{6}$O$_{13}$ measured under a zero magnetic field below 15 K. The inset shows the behavior in the whole temperature range. In these figures, the contribution from lattice vibrations has not been subtracted.}
\label{fig3}
\end{figure}

Measurements of specific heat (Fig.~\ref{fig3}) reveal two relatively broad local maxima at 7 K and 1 K. These features are also dissimilar to the sharp peaks observed in association with phase transitions accompanied by the development of long-range order. It can therefore be concluded that Co$_{4}$B$_{6}$O$_{13}$ is paramagnetic down to at least 0.4 K. 

\begin{figure}[tb]
\begin{center}
\includegraphics[width=1.0\linewidth]{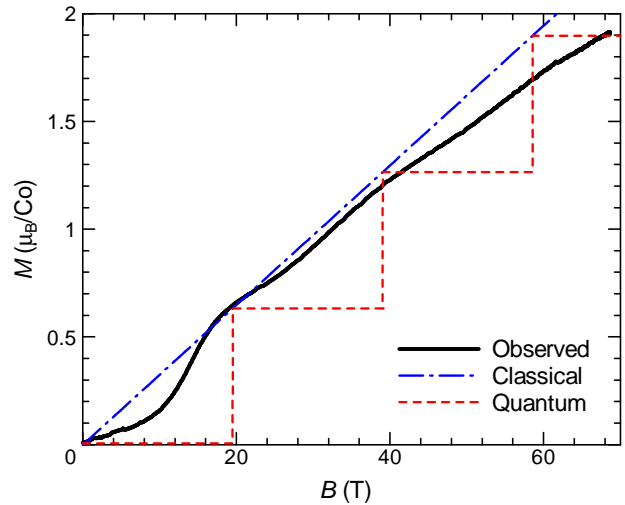}
\end{center}
\caption{Observed magnetization at 1.3 K under a magnetic field applied parallel to the [111] direction using a pulse magnet (solid line). Calculation based on a classical Heisenberg model and that on a quantum ($S=3/2$) Heisenberg model are also shown with dash-dotted line and dashed line, respectively. (See Eq.~(\ref{eq1})) In the calculation, we assume $J/k_{\mathrm{B}}=-33.2$ K and $g=2.53$ in accordance with the detaild calculation based on an exact diagonalization method.}
\label{fig4}
\end{figure}

The magnetization curve for Co$_{4}$B$_{6}$O$_{13}$ at 1.3 K under an increasing magnetic field of 0-70 T is shown in Fig.~\ref{fig4}. The magnetization at 70 T is close to 8 $\mu_{\mathrm{B}}$ per Co$_{4}$ cluster, equivalent to approximately half the saturation magnetization assuming $g = 2.53$. The relationship is characterized by a sharp increase in magnetization at 10-18 T, and a periodic undulation of the relationship at higher fields.

As a first step in interpreting this peculiar magnetization behavior, a simple model of a spin tetrahedron is considered.\cite{schnack06,dai04} The Hamiltonian of such a system is written as
\begin{eqnarray}
\label{eq1}
{\cal H} &=& -J\sum_{i<j} \mbox{\boldmath$s$}_{i} \cdot \mbox{\boldmath$s$}_{j}  - g\mu_\mathrm{B}\mbox{\boldmath$B$} \cdot \sum_{i} \mbox{\boldmath$s$}_{i}\nonumber\\
&=& -\frac{J}{2}\left( \mbox{\boldmath$S$}^{2} - \sum_{i} \mbox{\boldmath$s$}_{i}^{2} \right) - g\mu_\mathrm{B}\mbox{\boldmath$B$} \cdot \mbox{\boldmath$S$},
\end{eqnarray}
where $\mbox{\boldmath$s$}_{i}$ ($i = 1, 2, 3, 4$) denotes the spin of the $i$th atom forming the tetrahedron, $\mbox{\boldmath$S$}$ denotes the total spin ($= \mbox{\boldmath$s$}_{1} + \mbox{\boldmath$s$}_{2} + \mbox{\boldmath$s$}_{3} + \mbox{\boldmath$s$}_{4}$), $J$ is the exchange interaction, and $\mbox{\boldmath$B$}$ is the external magnetic field. In the antiferromagnetic case, $J$ is negative.

In the classical regime, the total magnetic moment per cluster is completely cancelled under a zero field, while a finite non-zero moment is induced under a small magnetic field because the spins continuously change direction with respect to the magnetic field. This behavior results in the linear magnetization curve shown in Fig.~\ref{fig4}, which fails to reproduce the periodic undulation of the observed curve.

In the quantum model, $\mbox{\boldmath$s$}_{i}$ should be treated not as a classical vector but as an operator. The energy eigenvalues of the Hamiltonian (Eq.~(1)) are easily obtained by
\begin{eqnarray}
\label{eq2}
E=-\frac{J}{2}\left\{ S(S+1)-4s(s+1) \right\} - g\mu_\mathrm{B}BS_{z},
\end{eqnarray}
where $S$ and $s$ denote the quantum numbers corresponding to the magnitude of the total spin per cluster and of each spin, respectively, $B$ is the magnetic field, and $S_{z}$ is the spin component of the total spin along the magnetic field. The total spin ($S$) is an integer from 0 to $4s$, and $S_{z}$ is an integer from $-S$ to $S$. This results in a staircase-like magnetization curve as shown in Fig.~\ref{fig4}, where the step width and step height are given by $|J|/(g\mu_{\mathrm{B}})$ and $g\mu_{\mathrm{B}}$, respectively. Such a magnetization curve qualitatively explains the undulating feature of the observed curve, although the observed relationship is considerably more smeared.

\begin{figure}[tb]
\begin{center}
\includegraphics[width=1.0\linewidth]{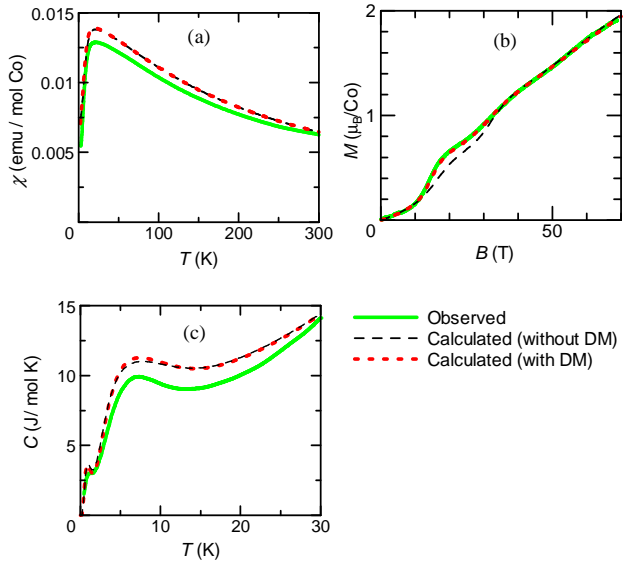}
\end{center}
\caption{Comparison between observed results and the calculations based on the exact diagonalization method. (Eq.~(\ref{eq3})) The fitted parameters are $J/k_{\mathrm{B}} = -33.2$ K, $D/k_{\mathrm{B}} = -19.3$ K, $g_{\parallel} = 2.53$, and $g_{\perp} = 2.43$. Dashed (dotted) lines correspond to the calculation without (inclusive of) the DM interaction $d$, where $d/k_{\mathrm{B}} = 0.68$ K. In the calculation of the specific heats (c), we added 1.5$\times10^{-4} \cdot T^{3}$ term as a lattice contribution.}
\label{fig5}
\end{figure}

For further clarification of this relationship, and to clarify the origin of the double-peak structure of the specific heat relationship and the non-zero susceptibility at low temperature, the simple model can be extended to include the interactions originally omitted in Eq. (1). The Hamiltonian for this expanded model is given by 
\begin{eqnarray}
\label{eq3}
{\cal H} &=& -J\sum_{<ij>} \mbox{\boldmath$s$}_{i} \cdot \mbox{\boldmath$s$}_{j} + D \sum_{i} \left( \mbox{\boldmath$s$}_{i} \cdot \mbox{\boldmath$e$}_{i} \right) ^{2}\nonumber\\
&& + \sum_{i<j}\mbox{\boldmath$d$}_{ij} \cdot \left( \mbox{\boldmath$s$}_{i} \times \mbox{\boldmath$s$}_{j} \right) - \mu_\mathrm{B} \sum_{i} { \mbox{\boldmath$B$} \widetilde{g}_{i} \mbox{\boldmath$s$}_{i} }.
\end{eqnarray}
In this Hamiltonian, the second term accounts for single-ion spin anisotropies, where $D > 0$ ($D < 0$) for easy-plane (easy-axis) anisotropy. The vector $\mbox{\boldmath$e$}_{i}$ characterising the anisotropy of the $i$th spin, is the unit vector parallel to the bond connecting the center of the tetrahedron to the position of the $i$th spin. The third term accounts for Dzyaloshinsky-Moriya (DM) antisymmetric interactions.\cite{elhajal05} For tetrahedral symmetry, it is explicitly given as $\mbox{\boldmath$d$}_{ij} = d(\mbox{\boldmath$e$}_{i} \times \mbox{\boldmath$e$}_{j}) / | \mbox{\boldmath$e$}_{i} \times \mbox{\boldmath$e$}_{j} | $. The final term is the Zeeman term, which takes into account the anisotropy of the $g$ factors for each spin, where $\widetilde{g}_{i}$ is a $g$-factor tensor having principal values of $g_{\parallel}$ parallel to $\mbox{\boldmath$e$}_{i}$ and principal values of $g_{\perp}$ perpendicular to $\mbox{\boldmath$e$}_{i}$. The eigenvalues can now be calculated by exact diagonalization of the spin Hamiltonian, optimizing the parameters $J$, $D$, $d$, $g_{\parallel}$ and $g_{\perp}$ so as to obtain the best fit to the observations. Firstly, we carried out the optimization without the DM term. The best-fit values are $J/k_{\mathrm{B}} = -33.2$ K, $D/k_{\mathrm{B}} = -19.3$ K, $g_{\parallel} = 2.53$, and $g_{\perp} = 2.43$. The calculated fits, shown in Fig.~\ref{fig5} as dashed lines, almost reproduce the behavior of the susceptibility and the specific heat. On the other hand, there is a discrepancy in the magnetization curves between 15 T and 35 T. Introduction of the DM interaction as large as $d/k_{\mathrm{B}} = 0.68$ K excellently improves the fitting, as shown with dotted lines in Fig.~\ref{fig5}.

\begin{figure}[tb]
\begin{center}
\includegraphics[width=1.0\linewidth]{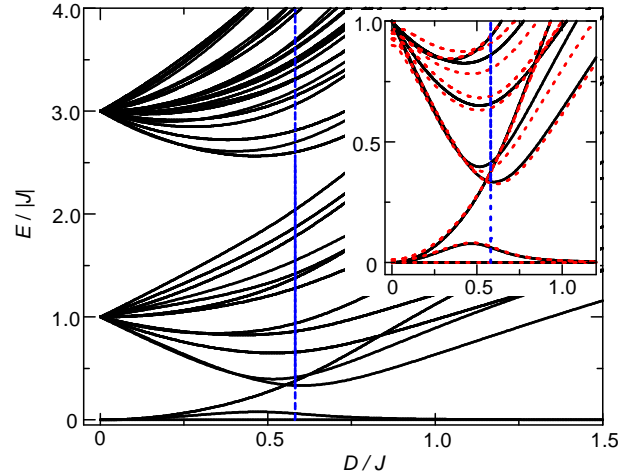}
\end{center}
\caption{Calculated energy diagram for a spin tetrahedron composed of $s = 3/2$ spins with respect to spin anisotropy ($D$). Calculated results are those determined by exact diagonalization of the spin Hamiltonian. The inset shows the detail of the low-energy region. Solid (dotted) lines denote calculations without (inclusive of) the DM interaction $d$. ($d/k_{\mathrm{B}} = 0.68$ K) Vertical line denotes the optimised ratio of $D/J$ affording the best fit to the experimental results.}
\label{fig6}
\end{figure}

The anisotropy ($D$) is the important factor in this optimization. At $D = 0$, a large number of eigenvalues are bundled at discrete levels, each corresponding to the eigenstate of the total spin number ($S$) in the quantum Heisenberg model (Eq.~(\ref{eq2})). When a finite $D$ is introduced, degenerate levels are split as a result of mixing between different $S$ states as shown in Fig.~\ref{fig6}. This results in the non-zero susceptibility at low temperature and the smearing of the ideal staircase-like magnetization curve. A group-theoretical analysis reveals that the ground state is four-fold degenerate for $S = 0$, splitting into three energy levels (the highest still doubly degenerate) with increasing $D$. This behavior gives rise to the double peak structure in the specific heat curve, where the peak at 1 K corresponds to the excitation between different types of singlet states. In Co$_{4}$B$_{6}$O$_{13}$, the single-ion anisotropy ($D$) lifts the degeneracy of the ground state, and may explain why the spin tetrahedra in this material are free from Jahn-Teller type distortion. This is in contrast to the pyrochlore lattice antiferromagnet such as CdCr$_{2}$O$_{4}$, \emph{etc.}, in which multi-fold degeneracy of the ground state is relieved by spin Jahn-Teller distortion.

The present results demonstrate that the coexistence of different types of singlet states is an important characteristic of spin tetrahedra. In the simplest $s = 1/2$ spin tetrahedron, the singlet states have two-fold degeneracy, assigned in one previous study to a dimer singlet and a plaquette singlet,\cite{koga01} and in another to two bases with opposite chirality.\cite{tsunetsugu01} It is interesting to remember that such degenerate nature of spin tetrahedron may make the ground state of the pyrochlore magnet exotic.

\section{Conclusion}

Through investigation of the magnetic properties of undistorted spin tetrahedra in Co$_{4}$B$_{6}$O$_{13}$, this compound was demonstrated to have an almost spinless ground state and a periodically undulating magnetization curve indicative of quantization of the total spin per spin tetrahedron. The specific heat exhibited two peaks at low temperature, which are interpreted as being due to a complex ground state composed of multiply degenerate singlet states reflecting the highly symmetric structure of the regular tetrahedron. In the present material, it would also be interesting to investigate magnetism under high pressure, which modifies the magnitude of the intercluster interactions and might cause a quantum phase transition to a long-range ordered state.

\section*{Acknowledgment}
The authors express their gratitude to Y. Ishii and H. Nakano for discussions. This study was supported in part by a Grant-in-Aid for Scientific Research on priority areas ``High Field Spin Science in 100 T'' (No. 451) from the Ministry of Education, Culture, Sports, Science and Technology (MEXT) and by a Grant-in-Aid for Scientific Research from the Japan Society for the Promotion of Science.

\end{document}